# Characteristic Lengths of Interlayer Charge-Transfer in Correlated Oxide Heterostructures

*Ganesh Ji Omar,*[†,#] *Mengsha Li,*[‡] *Xiao Chi,*[§] *Zhen Huang,*[*,#] *Zhi Shiuh Lim,*[†,#] *Saurav Prakash,*[#,∇] *Shengwei Zeng,*[†,#] *Changjian Li,*[‡] *Xiaojiang Yu,*[§] *Chunhua Tang,*[‡] *Dongsheng Song,*[#,¶] *Andrivo Rusydi,*[†,§] *Thirumalai Venkatesan,*[†,‡,#,∇,○] *Stephen John Pennycook,*[‡] *Ariando Ariando*[*,†,#,∇]

[†]Department of Physics, National University of Singapore, Singapore 117542, Singapore

[‡]Department of Materials Science and Engineering, National University of Singapore, Singapore 117575, Singapore

[§]Singapore Synchrotron Light Source, National University of Singapore, 5 Research Link, Singapore 117603, Singapore

[#]NUSNNI-NanoCore, National University of Singapore, Singapore 117411, Singapore

[∇]National University of Singapore Graduate School for Integrative Sciences and Engineering (NGS), 28 Medical Drive, Singapore 117456, Singapore

[○]Department of Electrical and Computer Engineering, National University of Singapore, Singapore 117576, Singapore

[¶]Present address: Ernst Ruska-Centre for Microscopy and Spectroscopy with Electrons. Forschungszentrum Jülich, D-52425 Jülich, Germany

*ariando@nus.edu.sg ; nnihz@nus.edu.sg

ABSTRACT

Using interlayer interaction to control functional heterostructures with atomic-scale designs has become one of the most effective interface-engineering strategies nowadays. Here, we demonstrate the effect of a crystalline $LaFeO_3$ buffer layer on amorphous and crystalline $LaAlO_3/SrTiO_3$ heterostructures. The $LaFeO_3$ buffer layer acts as an energetically favored electron acceptor in both $LaAlO_3/SrTiO_3$ systems, resulting in modulation of interfacial carrier density and hence metal-to-insulator transition. For amorphous and crystalline $LaAlO_3/SrTiO_3$ heterostructures, the metal-to-insulator transition is found when the $LaFeO_3$ layer thickness crosses 3 and 6 unit cells, respectively. Such different critical $LaFeO_3$ thicknesses are explained in terms of distinct characteristic lengths of the redox-reaction-mediated and polar-catastrophe-dominated charge transfer, controlled by the interfacial atomic contact and Thomas-Fermi screening effect, respectively. Our results not only shed light on the complex interlayer charge transfer across oxide heterostructures but also provides a new route to precisely tailor the charge-transfer process at a functional interface.

KEYWORDS: interface engineering, buffer layer, perovskite oxide heterostructures, charge transfer.

INTRODUCTION

Charge transfer has been one of the most intriguing phenomena responsible for rich physical properties in condensed matter.[1-3] For instance, the electron transfer from the doped AlGaAs to the adjacent undoped GaAs layer results in the formation of the high-mobility two-dimensional electron gas at the well-known GaAs/AlGaAs heterostructure.[4] Another example is the unconventional orbital reconstruction at the YBa$_2$Cu$_3$O$_7$/(La,Ca)MnO$_3$ interface, wherein holes get transferred from YBa$_2$Cu$_3$O$_7$ to (La,Ca)MnO$_3$.[5,6] Charge transfer (or tunneling) across a multiferroic layer has been utilized to realize multiple switchable states in an all-oxide-based memory device.[7-9] Recent studies on two-dimensional materials have also highlighted the crucial role of charge transfer in photocatalysis[10,11] and water splitting.[12] The physical concept of charge transfer is essential for both the fundamental understanding of interfacial properties and the future design of novel functional devices.

The complexity of interlayer charge transfer is further evidenced in the case of LaAlO$_3$/SrTiO$_3$ (LAO/STO) heterostructures,[13,14] which host rich and exotic emergent properties.[15-18] It is believed that the formation of the two-dimensional electronic system (2DES) at the LAO/STO interface is ascribed to the electron transfer from the LAO layer to the STO substrate surface.[14,19,20] Among various models that have been proposed to explain the origin of charge transfer in the last 15 years, there are two most popular mechanisms, i.e, *polar catastrophe*[14] and *redox reaction*.[21,22] In the model of polar catastrophe, the charge transfer occurs to compensate for the discontinuity of formal polarization between the polar crystalline LAO layer and nonpolar TiO$_2$-terminated STO substrate, leading to the formation of a 2DES at the crystalline LAO/STO (*c*-LAO/STO) interface.[13,14] On the other hand, due to the different oxygen affinity between LAO and STO, the oxygen-deficient LAO layer will extract oxygen from the adjacent STO layer, leaving oxygen vacancies and mobile electrons at the interface. The latter explains the appearance of the 2DES at the amorphous LAO/STO (*a*-LAO/STO) interface with no polar catastrophe.[21,23] Furthermore, other factors such as chemical doping,[24,25] off-stoichiometry,[26,27] intermixing[28,29], and surface defects[30,31] can also affect the interlayer charge transfer in the LAO/STO system.

Although the 2DES induced by various charge-transfer mechanisms may share some similar physical properties, different scenarios must be considered to understand the *c*-LAO/STO and *a*-LAO/STO system. Specifically, the polar-catastrophe-dominated charge transfer at the *c*-LAO/STO interface can be described by electrostatic screening – the transferred electrons are distributed in the interfacial STO layer to screen the polar field in LAO. If a thin buffer layer is inserted to serve as an energetically-favored electron acceptor, the electron transfer between LAO and STO will be suppressed, or even completely blocked when the buffer layer thickness exceeds the Thomas-Fermi screening length $\lambda$, which can be viewed as the characteristic length of the polar-catastrophe-dominated charge transfer. On the other hand, redox-reaction-mediated charge transfer will be interrupted by inserting an inert buffer layer which prevents direct contact between the amorphous LAO layer and the STO substrate surface. Considering the thickness of the interdiffusion layer [1-2 unit cells (uc)] induced by the growth process and substrate atomic terrace (Figure S8, Supporting Information), a 2(3) uc thick inert buffer layer will block most (all) of the atomic contact between LAO and STO, and thus prevent the redox-reaction-mediated charge transfer. This is consistent with the previous observation that the 2 uc inert (La,Sr)MnO$_3$ layer is thick enough to turn the conducting *a*-LAO/STO interface into the insulating state.[32] Hence, we argue that the universal characteristic length of redox-reaction-mediated charge transfer should be around 2-3 uc, at which thickness of the inert buffer layer is able to separate the *a*-LAO and STO to avoid interlayer charge transfer.

RESULTS

In this paper, we use a crystalline LaFeO$_3$ (LFO) as a buffer layer to investigate the interlayer charge transfer in the *a*- and *c*-LAO/STO systems, as sketched in the inset of Figure 1a and 1b. There are three reasons for adopting the LFO buffer layer. First, the layer-by-layer growth of LFO on the STO substrate (Figure S1, Supporting Information) enables us to control the buffer layer thickness down to 1 uc (~ 0.4 nm) to precisely examine the characteristic length of the charge transfer. Second, the bandgap of LFO (2.34 eV)[33] is smaller than that of STO (3.2 eV)[34] and LAO (5.6 eV)[35] which means that the LFO can serve as an energetically-favoured electron acceptor compared to STO during the interlayer charge transfer.[36] Third, there is neither polar-catastrophe-dominated nor redox-reaction-mediated charge transfers across the LFO/STO interface when the LFO layer is thin. Based on the polar-catastrophe model, the critical thickness of LFO ($l_{\mathrm{LFO}}$) for triggering the interlayer charge transfer between polar LFO and non-polar STO is about 20 nm (50 uc), which can be estimated by $l_{\mathrm{LFO}} = \varepsilon_0 \varepsilon_{\mathrm{LFO}} \Delta E_g / e P^0_{\mathrm{LFO}}$,[37] where $\varepsilon_0$ is the vacuum permittivity, $\varepsilon_{\mathrm{LFO}} = 120$[38,39] is the relative permittivity of LFO, $e$ is the elementary charge, $P^0_{\mathrm{LFO}}$ is the formal polarization of LFO and $\Delta E_g$ (~ 3.2 eV) is the energy difference between the valence band of LFO and the conduction band of STO. In the view of oxygen affinity, LFO cannot take the oxygen from STO to induce the interfacial redox reaction, resulting in preservation of insulating nature at the STO interface.[40] Therefore, when the LFO layer thickness is below $l_{\mathrm{LFO}}$, neither polar-catastrophe-dominated nor redox-reaction-mediated charge transfer occurs at the LFO/STO interface.

Figure 1a and 1b show the 3 K sheet conductance $G_{\mathrm{Sheet}}$ of the *a*- and *c*-LAO/LFO/STO interfaces, respectively, as a function of LFO buffer layer thickness *t*. Here, the *a*-LAO layer thickness is fixed at 3 nm and the *c*-LAO layer at 8 uc for all samples. The *a*-LAO/LFO/STO samples exhibit a metal-to-insulator transition (MIT) when $t$ increases from 2 to 3 uc. This is consistent with our previous discussion on the redox-reaction-mediated charge transfer, where the characteristic length is 2-3 uc. By contrast, the characteristic length of the $c$-LAO/LFO/STO interface is 6 uc, above which the LFO buffer layer turns the $c$-LAO/LFO/STO heterostructure into the insulating state. The observed MIT might be ascribed to either the low carrier density $n_S$ or the suppressed carrier mobility $\mu$ of the 2DES, corresponding to the modulation doping or disorder scattering induced by the LFO buffer layer, respectively. Figure 1c and 1d show the gradual drop of $n_S$ with increasing $t$ in both $a$- and $c$-LAO/LFO/STO samples. Meanwhile, Figure 1e reveals that decreasing $n_S$ enhances the $\mu$ of the 2DES (Figure S2, Supporting Information), suggesting the modulation doping, rather than the disorder scattering, plays a dominant role in the LFO-buffered $a$- and $c$-LAO/STO heterostructures.

Despite the difference in the characteristic length and critical LFO buffer layer thickness, the conducting $a$- and $c$-LAO/LFO/STO interfaces with a thin LFO buffer layer ($t \leq 2$ uc) show a similar temperature-dependent carrier density. As illustrated in Figure 2a and 2b, when $t$ is 0 or 1 uc, an obvious drop in carrier densities is observed when the temperature is decreased from 100 to 10 K in both $a$- and $c$-LAO/LFO/STO samples. The activation energy $\Delta E$, corresponding to the energy gap between the trapping centers and conduction band minimum as sketched in the inset of Figure 2(c), can be estimated by $n_S \propto e^{(-\Delta E/(k_B T))}$ where $k_B$ is the Boltzmann constant and $T$ is the temperature (Figure S3, Supporting Information). Accordingly, the calculated $\Delta E$ for $t = 0$ and 1 uc is around 2-5 meV, which is consistent with the results reported in the oxygen-deficient STO-based 2DES with oxygen vacancies as the trapping centers.[23] When $t$ is increased to 2 uc, both the $a$- and $c$-LAO/LFO/STO samples show nearly degenerate carrier densities ($\Delta E \sim$ 0-1 meV). This degenerate conducting behavior has also been reported when the formation of oxygen vacancies is suppressed in STO-based interfaces.[41] Further increasing the LFO layer thickness to 3 uc, the *a*-LAO/LFO/STO interfaces become completely insulating while the *c*-LAO/LFO/STO interfaces still maintain the metallic state with degenerate carriers. However, a large enhancement of $\Delta E$ ~ 40-50 meV is observed in the metallic *c*-LAO/LFO/STO samples with *t* = 4-5 uc. Such a large variation of $\Delta E$, as summarized in Figure 2c, suggests a new type of trapping center that replaces the oxygen vacancies at the *c*-LAO/LFO/STO interface with *t* = 4-5 uc. In other words, for *c*-LAO/LFO/STO samples, the LFO buffer layer will suppress firstly the redox-reaction-mediated and then the polar-catastrophe-dominated interlayer charge transfer between the LAO and STO.

DISCUSSION

To better understand the role of the LFO buffer layer in the charge-transfer process, atomically-resolved scanning transmission electron microscopy (STEM) accompanied by electron energy-loss spectroscopy (EELS) were performed on the *c*-LAO/LFO/STO sample with *t* = 4 uc. As seen in Figure 3a, the interdiffusion is confined within the 1-uc-thick interfacial layer, resulting in sharp interfaces across the heterostructure. Specifically, the sublayer structure at the LFO/STO and LAO/LFO interfaces are LaO/TiO$_2$ and LaO/FeO$_2$, respectively. It confirms the BO$_2$-terminated surface is maintained on the SrTiO$_3$ substrate, as well as the LFO and LAO layers. Figure 3b displays the evolution of the layer-resolved EELS spectra around the Fe-$L_{2,3}$ edge obtained from the 4-uc-thick LFO buffer layer. When compared to the bottom LFO layer (close to the LFO/STO interface), the top LFO layer (close to the LAO/LFO interface) shows enhanced intensity at the low-energy part around both $L_3$ and $L_2$ peaks. This indicates a relatively larger amount of Fe$^{2+}$ ions are located close to the LAO/LFO interface. By fitting the Fe-$L_{2,3}$ edge by a combination of Gaussian and Lorentzian functions, the intensity ratio between $L_2$ and $L_3$ was calculated to obtain the $Fe^{2+}$ concentration, denoted by $Fe^{2+}/(Fe^{2+}+Fe^{3+})$. The $Fe^{2+}$ fraction is 24±5% in the top LFO layer and reduces close to 0% for the rest of the LFO layers as shown in supporting information Table S1. The $Fe^{2+}$ ions can be formed in the LFO layer due to either the oxygen-deficient growth condition or interlayer charge transfer. In order to clarify the origin of $Fe^{2+}$, X-ray absorption spectroscopy (XAS) is applied on various control samples, including the *a*-LAO/LFO/STO, *c*-LAO/LFO/STO and LFO/Nb-SrTiO$_3$ samples, with the same thickness (*t* = 1 uc) and growth parameters (oxygen pressure at 10 mTorr) for the LFO layer. In our proposed model, interlayer charge transfer is expected to occur in the *a*- and *c*-LAO/LFO/STO samples, but not in the LFO/Nb-SrTiO$_3$. In Figure 3c, when compared to the non-charge-transfer LFO/Nb-SrTiO$_3$ sample, the charge-transfer *a*- and *c*-LAO/LFO/STO interfaces exhibit an obvious enhancement of spectral weight below the $Fe^{3+}$ $L_3$ main peak (709 eV).[42,43] This suggests that the main origin of the $Fe^{2+}$ formation in the LFO layer must be ascribed to the interlayer charge transfer mediated by the interfacial polar catastrophe or redox reaction, instead of the formation of oxygen vacancies induced by the vacuum environment during the film growth. Also, by linearly fitting the XAS data of the *c*-LAO/LFO/STO interface using the theoretical $Fe^{2+}$ and $Fe^{3+}$ data,[44] the concentration of $Fe^{2+}$ ions that are confined within the 1-uc-thick LFO buffer layer is around 18±5% (Figure S4, Supporting Information), which is in line with our EELS results. Our observation of $Fe^{2+}$ ions located close to the LAO/LFO interface supports our argument that the LFO acts as an energetically favored electron acceptor in the interlayer charge transfer.

However, those $Fe^{2+}$ ions are not energetically favored at the B-site in ABO$_3$ perovskite lattices due to a large ionic radius. So, the observed B-site $Fe^{2+}$ ions are only stable at the interface, and thus well confined within 1-uc layer of LFO at the LAO/LFO interface (see Table S1, Supporting Information). This phenomenon is quite different from the LAO/STO interface, where both $Ti^{3+}$ and $Ti^{4+}$ are suitable for B-site of the perovskite. And the conductivity at the LAO/STO interface is ascribed to the $Ti^{3+}$ ions that propagate several unit cells (even nm) from the interface into the STO substrate – because the $Ti^{3+}$ ions close to the interface are localized due to Anderson localization. So, in the case of LAO/LFO/STO, the transfer charges (or $Fe^{2+}$ ions) in the LFO layer are localized and do not contribute to the conductivity of 2DES.

Now we focus on explaining why the LFO-induced MIT occurs at $t = 6$ uc in the *c*-LAO/LFO/STO samples. As sketched in Figure 4a, for the conventional *c*-LAO/STO with $t = 0$ uc, the negatively charged electrons are transferred from LAO to STO to compensate for the built-in potential in the polar LAO layer with alternative stacking of $LaO^+$ and $AlO^-$ sublayers. Most of the transferred electrons are confined in the interfacial STO layer with thickness $\sim \lambda_{STO}$, corresponding to the Thomas-Fermi screening length of STO. When the LFO buffer layer is inserted with $t < \lambda_{LFO}$, as plotted in Figure 4b, some of the transferred electrons are trapped by the low-energy state provided by the LFO buffer layer due to its small bandgap.[36] The negatively charged LFO buffer layer, accompanied by the formation of $Fe^{2+}$, can also lower the built-in potential in the LAO layer to avoid the polar catastrophe. On the other hand, the rest of the transferred electrons are still able to penetrate through the LFO buffer layer to form the 2DES but with the lower charge density at the STO side. Further increasing $t$ to $\lambda_{LFO}$ or above, all of the transferred electrons are trapped by the LFO buffer layer. In other words, the polar field of the crystalline LAO is completely screened by the LFO buffer layer with $t > \lambda_{LFO}$, and there is no electron transferred to the STO. Accordingly, the MIT occurs and no 2DES is formed at the STO as shown in Figure 4c. Based on the Thomas-Fermi screening model, $\lambda_{LFO}$ can be calculated by $\lambda_{LFO} = \sqrt{\frac{\varepsilon_0 \varepsilon_{LFO} k_B T}{n_c e^2}}$, where $n_c$ is the charge density in LFO. Given that LFO is insulating in our samples, $n_c$ can be estimated by the Mott criterion[45] with $n_c \approx \left(0.25/a^*\right)^3$, where $a^* = 4\pi\varepsilon_{LFO}\varepsilon_0\hbar^2/m^*_{LFO}e^2$ is the effective Bohr radius with $m^*_{LFO}$ is effective mass of electrons in the LFO layer. Assuming $m^*_{LFO} \approx 2m_0$,[46] the estimated $n_c$ will be ~ $3.5 \times 10^{17}$ cm$^{-3}$, which is close to the density of transfer charges between LFO and 2DES (~ 4-5 × $10^{17}$ cm$^{-3}$ in Figure S9 and S10, Supporting Information). Accordingly, the estimated value of $\lambda_{LFO}$ is 2.4 nm (or 6 uc). This is quantitatively consistent with our experimental observation that the MIT occurs at $t$ = 6 uc in the $c$-LAO/LFO/STO samples.

CONCLUSION

In summary, we systematically studied the $a$- and $c$-LAO/STO interfaces by using a crystalline LFO buffer layer. The LFO buffer layer provides a low-energy state to capture the transferred electrons. For the $a$- ($c$-) LAO/LFO/STO interfaces, a 3-uc-thick (6-uc-thick) LFO layer can completely block the interlayer charge transfer mediated by redox reaction (polar catastrophe), leading to the MITs observed in our samples. The different critical LFO layer thicknesses correspond to the different charge-transfer characteristic lengths crossing the $a$- and $c$-LAO/LFO/STO heterostructures. For the $a$-LAO/LFO/STO interface, the characteristic length is determined by the interfacial atomic contact between Al ions and the TiO$_2$ sublayer; while for the $c$-LAO/LFO/STO, the characteristic length can be expressed by Thomas-Fermi screening length. Our data reveal the complexity of interlayer charge transfers crossing the oxide heterostructure, and also provides a buffer-layer-based strategy for engineering the charge-transfer process in correlated oxide heterostructures.

EXPERIMENTAL SECTION

*Sample Preparation:* The 0.5-mm-thick (001) SrTiO$_3$ (STO) substrate (Crystec) was treated with HF and annealed to obtain defined terrace steps and TiO$_2$-terminated surface. The pulsed laser deposition method was used for sample preparation. The layer-by-layer growth of LaFeO$_3$ (LFO) and crystalline LaAlO$_3$ (c-LAO) films was monitored by *in-situ* reflection high-energy electron diffraction (RHEED). The LFO layer thickness was changed from 0 to 8 uc, the *c*-LAO fixed at 8 uc, and the amorphous LAO (*a*-LAO) fixed at 3 nm. During the growth, a nanosecond KrF 248nm laser was used with a fluence of 1.5 Jcm$^{-2}$ and a repetition rate of 2 Hz. The LFO layer was deposited at temperature $T$ = 760 °C and oxygen partial pressure $P_{O2}$ = 10 mTorr. For the *c*-LAO layer, the parameters were $T$ = 740 °C and $P_{O2}$ = 0.5 mTorr, whereas *a*-LAO was grown at room temperature (RT) and same $P_{O2}$ = 0.5 mTorr. All the samples were cooled down to room temperature at the deposition P$_{O2}$ and at a rate of 10 °C/ min to 700 °C and 15 °C/ min to RT. Later, we also performed the thermal treatment by suitable annealing the *c*- and *a*-LAO/LFO/STO samples to remove the oxygen vacancies. A detailed elaboration of the thermal treatment can be found in Figure S6, supporting information. The surface and topography of the samples were examined by atomic force microscopy (AFM) using an Agilent 5500 system.

*Transport measurement:* The sheet resistance, carrier densities and carrier mobility are measured by Van der Pauw method on a 9 T Physical Property Measurement System (PPMS, Quantum Design).

*STEM Imaging and EELS Analysis:* TEM samples were prepared by a FIB (focused ion beam) system (FEI Versa 3D) with 30 kV Ga ions, followed by a low-voltage (i.e., 2 kV) cleaning step. STEM imaging was performed by a JEM-ARM200F (JEOL) microscope equipped with an ASCOR aberration corrector, a cold-field emission gun and a Gatan Quantum ER spectrometer, operated at 200 kV. The EELS results were acquired using a collection angle of 100 mrad with the energy dispersion of 0.25 and 0.1 eV per channel for the elemental mapping and energy loss near edge structure (ELNES), respectively. The EELS maps were subject to noise reduction by a principal component analysis (PCA) filter.

*X-Ray Absorption at the Fe $L_{3,2}$ and Ti $L_{3,2}$ edges:* XAS measurements were performed at the SINS (Surface, Interface and Nanostructure Science) beamline of the Singapore synchrotron light source. The energy resolution was set ~ 0.25 eV.

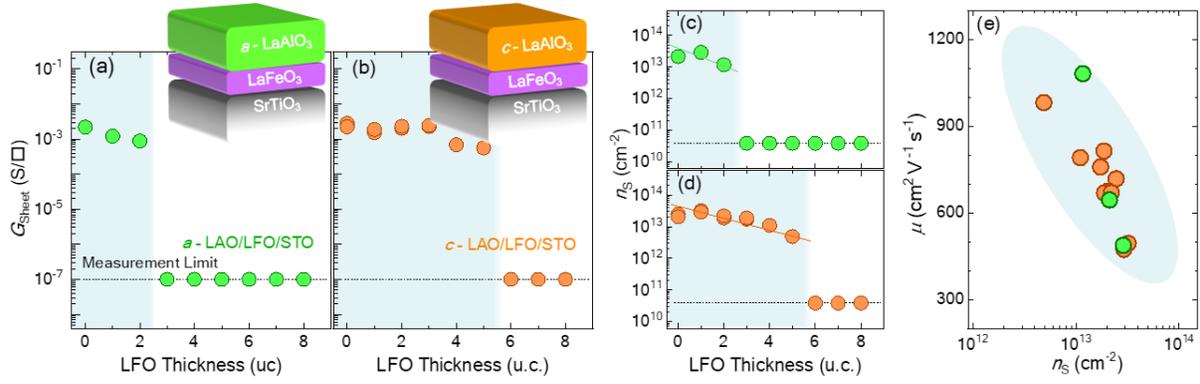

**Figure 1. Formation of the 2DES at the *a*- and *c*-LAO/LFO/STO heterointerface.** The LFO-thickness-dependent $G_{Sheet}$ (3 K) at *a*- and *c*-LAO/LFO/STO heterointerfaces is demonstrated in (a) and (b) with a schematic image of the LFO-buffered *a*- and *c*-LAO/STO (001) interface in the inset, while $n_S$ (3 K) in (c) and (d), respectively. (e) The relationship between $\mu$ and $n_S$ at 3 K for the LFO-buffered 2DES with modulation doping.

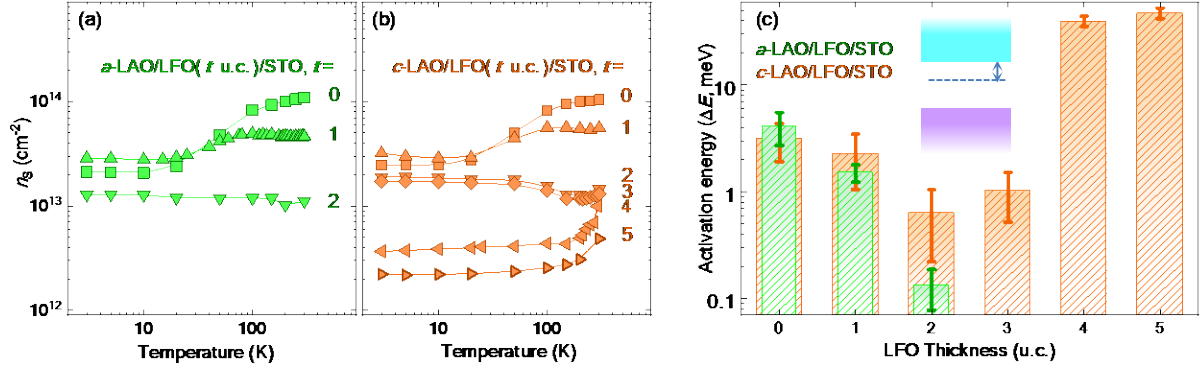

**Figure 2. Temperature dependence of transport properties in the LFO-buffered 2DES.** The temperature-dependent $n_S(T)$ for *a*- and *c*-LAO/LFO/STO samples with different *t* is summarized in (a) and (b), respectively. (c) The estimated thermal activation energies $\Delta E$ for various LFO-buffered 2DESs. Inset: Sketch of defining $\Delta E$ as indicating the energy gap from trapping centers to the conduction band minimum.

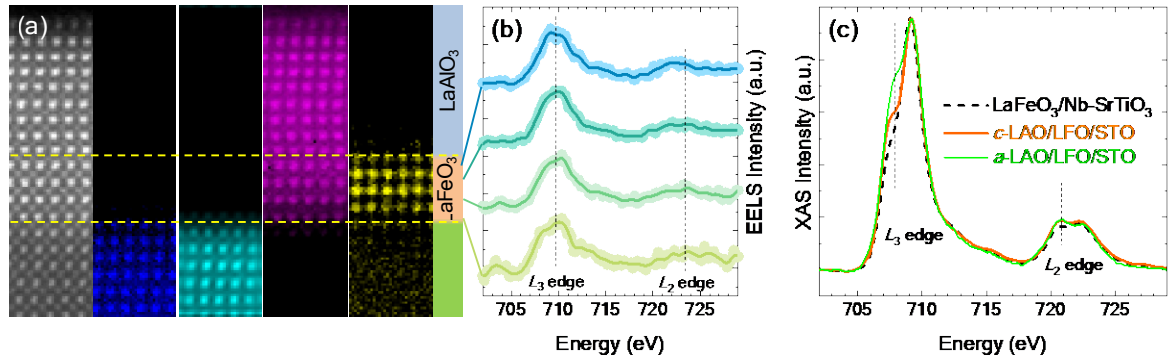

**Figure 3. Spectroscopic evidence of the Fe$^{2+}$ formation.** (a) Atomic structures obtained by STEM annular dark-field imaging (ADF, leftmost) and corresponding EELS mapping (Ti, Sr, La, and Fe, from left to right) in the *c*-LAO/LFO/STO heterostructure with the LFO layer thickness at 4 uc and LAO at 8 uc. (b) Plane-resolved EELS analysis of the Fe $L_{2,3}$ ELNES in the 4-uc-thick LFO buffer layer. (c) XAS results around Fe $L_{2,3}$ edges for the *a*-LAO/LFO/STO, *c*-LAO/LFO/STO and LFO/Nb-doped STO with the LFO layer thickness fixed at 1 uc. In (b) and (c), the Fe$^{2+}$-induced low-energy spectral weight enhancement is indicated by dashed lines.

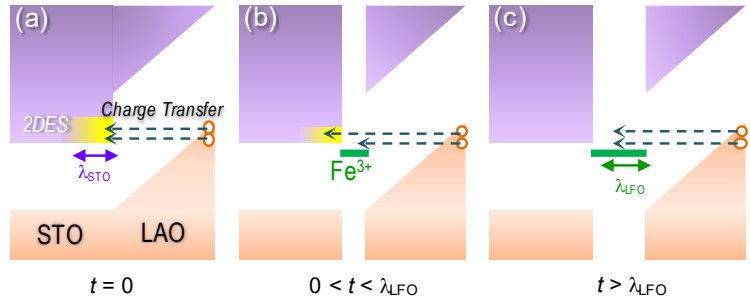

**Figure 4. Evolutions of the charge transfer across the *c*-LAO/LFO/STO with LFO layer thickness *t*.** (a) *t* = 0, when the depth distribution of the 2DES is close to the STO screening length $\lambda_{STO}$. (b) $0 < t < \lambda_{LFO}$, when transferred charges can still penetrate through the LFO layer and reach the STO side to form a 2DES with a reduced $n_S$. (c) $t > \lambda_{LFO}$, when transferred electrons are completely blocked and received by the LFO buffer layer.

**Author Contributions**

G.J.O., Z.H., and A.A. conceived the project idea. G.J.O. and M.S.L. contributed equally to this work. G.J.O. carried out the growth and transport measurement of the samples. M.S.L., C.L., and C.T. performed the TEM and EELS measurement and D.S.S. facilitated for simulation analysis. X.C. and X.Y. performed the XAS measurement. Z.H. assisted for interpretation of the theoretical model. Z.S.L., S.P., and S.W.Z. provided the suggestions and contributed to measurement and discussion. G.J.O., Z.H. and A.A. wrote the manuscript with help from other authors. All authors have given approval to the final version of the manuscript.

**Funding Sources**

This research is supported by the Agency for Science, Technology and Research (A*STAR) under its Advanced Manufacturing and Engineering (AME) Individual Research Grant (IRG) (A1983c0034), the National University of Singapore (NUS) Academic Research Fund (AcRF Tier 1 Grant No. R-144-000-391-144 and R-144-000-403-114) and the Singapore National Research Foundation (NRF) under the Competitive Research Programs (CRP Award No. NRF-CRP15-2015-01). C.J.L acknowledges the financial support from the Lee Kuan Yew Postdoctoral Fellowship through the Singapore Ministry of Education Academic Research Fund Tier 1 (R-284-000-158-114). S.J.P would like to acknowledge the financial support by the Ministry of Education, Singapore under its Tier 2 Grant (Grant No. MOE2017-T2-1-129).

**Notes**

The authors declare no competing financial interest.

The Supporting Information includes the details for the sample growth (Figure S1), temperature-dependent transport properties (Figure S2), estimation of activation energy (Figure S3), fitting of XAS/XLD/EELS (Figures S4-5 and Table S1), post-annealing effect (Figures S6-7), atomic contact at the heterointerface (Figure S8), and estimation of carrier density in LFO layer (Figure S9-10). This material is available free of charge via the internet at http://pubs.acs.org.

# SUPPORTING INFORMATION

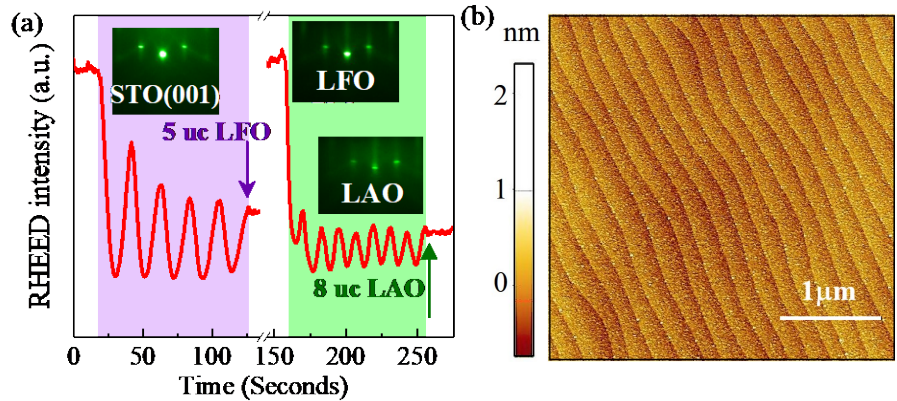

**Figure S1.** (a) Layer-by-layer growth of 8 uc LaAlO$_3$ and 5 uc LaFeO$_3$ on TiO$_2$ terminated SrTiO$_3$ interfaces determined by RHEED oscillations. The inset shows the clear streaky RHEED pattern reflections and diffraction peaks intensity along the (001) crystallographic direction during each growth, indicating high quality film deposition. Similar patterns were observed for all the samples in *c*-LAO/LFO/STO series. (b) The related surface morphology (3.5 μm × 3.5 μm) after the film growth confirmed by atomic force microscopy shows smooth regular terraces surface following the STO substrate.

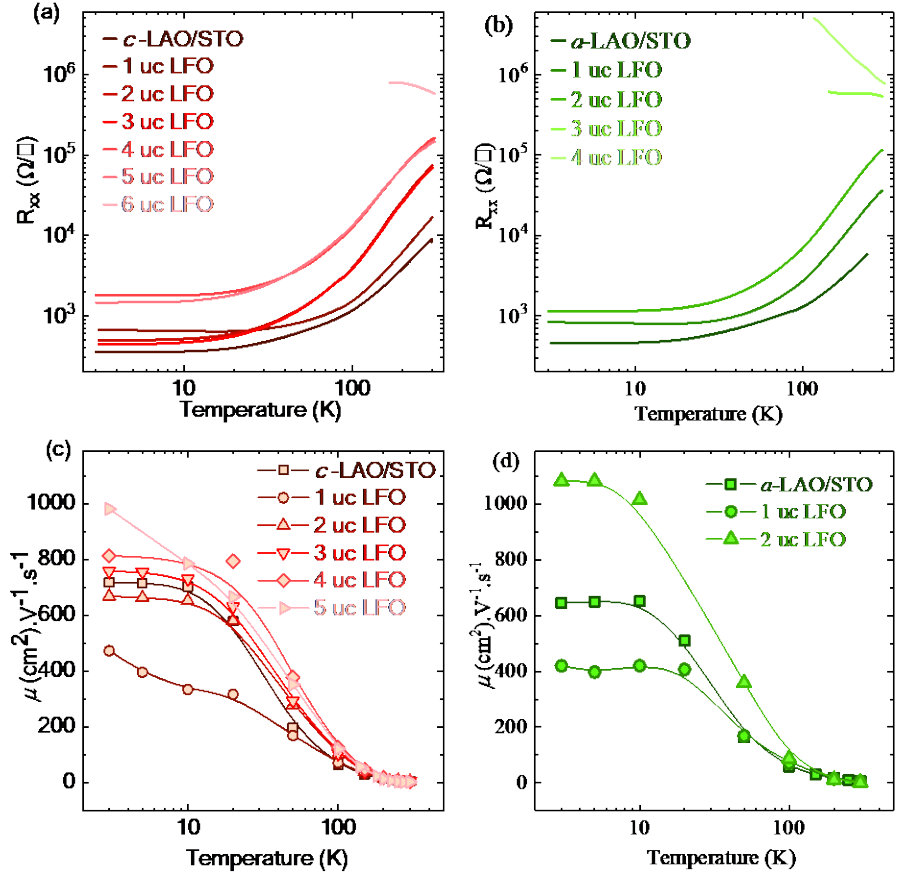

**Figure S2.** (a) and (b) Temperature dependence of sheet resistance ($R_{xx}$) for *a*- and *c*-LAO/LFO/STO with various LFO layer thicknesses. The *c*-LAO/LFO/STO sample preserves the metallicity up to the 5 uc LFO buffer layer, while it becomes insulating for the 6 uc LFO buffer, shown in red gradient colours. On the other hand, the *a*-LAO/LFO/STO sample is metallic up to 2 uc LFO and it shows insulating behaviour when the LFO layer is 3, 4 uc or above, indicated in green gradient colours. (c) and (d) Temperature dependence of mobility ($\mu$) for various thicknesses of LFO in *c*-LAO and *a*-LAO shown in red and green gradient colours respectively.

Many factors, such as stoichiometry, defects and intermixing, can affect the two-dimensional electron system (2DES) at the oxide interface. However, they cannot well explain the different electrical properties between crystalline and amorphous LAO/LFO/STO samples.

1) *Stoichiometry*: it is well known that the stoichiometry of a deposited film is mainly determined by the laser energy (1.5 Jcm$^{-2}$), target-substrate distance (7 cm) and growth pressure (10mTorr for LFO and 0.5mTorr for *a*-/ *c*-LAO) during the pulsed laser deposition process. Even for LAO/STO interface, those parameters have been used to control LAO's stoichiometry and thus transport properties. However, for our crystalline and amorphous samples, those growth parameters were kept unchanged to minimize the stoichiometry effect.

2) *Interfacial Defects & Intermixing*: It is true that the defects in the crystalline LAO are different from ones in the amorphous layer. However, given that the 2DES is hosted by the interfacial STO layer, the defects/intermixing that are closer to the interfacial STO layer shall play more significant effects on the 2DES. So, the defects/intermixing at the LFO/STO interface shall be more important than ones in the LAO layer. Because the LFO/STO interface is almost identical in both amorphous and crystalline LAO/LFO/STO samples, we believe the contribution from interfacial defects/intermixing is similar in all our samples.

More importantly, if the stoichiometry/defects/intermixing of amorphous and crystalline LAO layer is different and important, we are expecting to see different transport behaviors at the *c*- and *a*-LAO/LFO/STO with the same LFO layer thickness. However, we observe similar transport behaviors (temperature-dependent sheet resistances, carrier densities, etc.) at the *c*- and *a*-LAO/LFO/STO samples with $t_{LFO}$ = 1 and 2 uc. Those experimental facts suggest that the stoichiometry/defects/intermixing of amorphous and crystalline LAO layers should be similar, or at least not playing dominant roles in LFO-buffered LAO/STO systems.

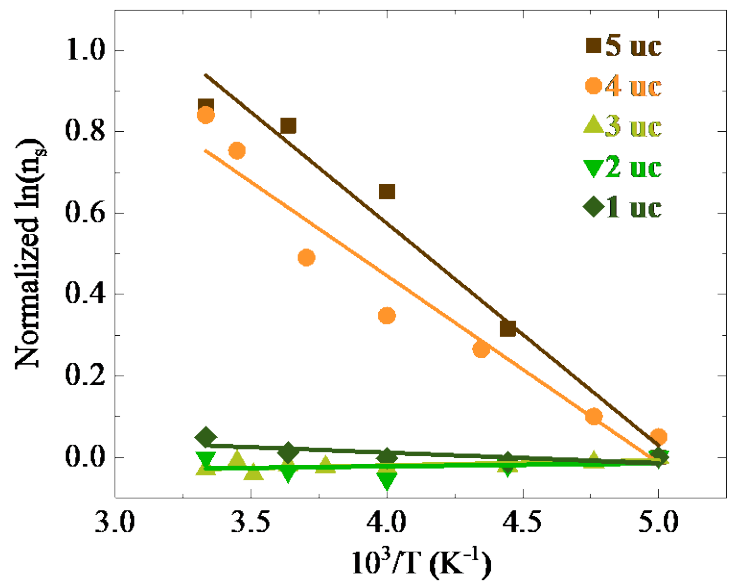

**Figure S3.** Arrhenius plot of ln($n_S$) vs 1000/T (1/K) for the temperature range 300-200 K and the linear fitting estimated from $n_S \propto e^{(-\Delta E/(k_B T))}$,[2] where $n_S$ is the sheet carrier density, $k_B$ is the Boltzmann constant, $T$ is the temperature and $\Delta E$ is the activation energy between the various trapping centers and conduction band minima.

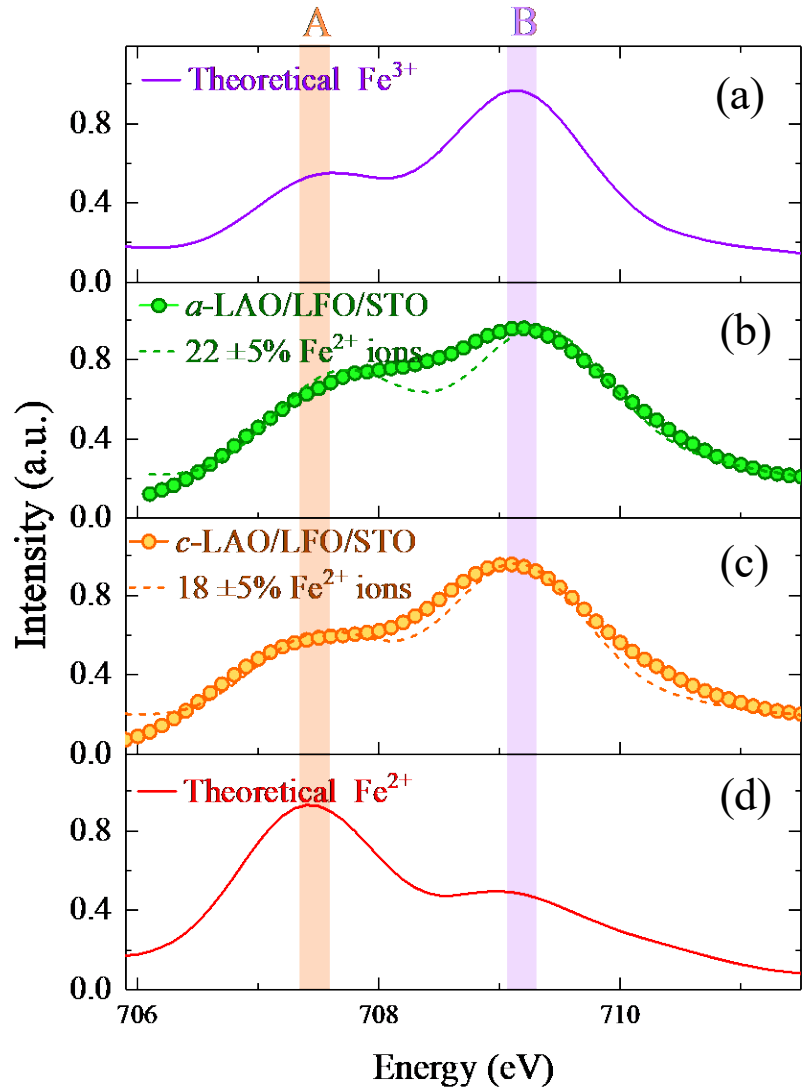

**Figure S4.** Linearly fitted XAS data of the LAO/LFO/STO interface using the theoretical $Fe^{3+}$ and $Fe^{2+}$ data as shown in (a) and (d).[3] The concentration of $Fe^{2+}$ ions that are confined within the 1-uc-thick LFO buffer layer is around (b) 22±5% for *a*-LAO/LFO/STO and (c) 18±5% for *c*-LAO/LFO/STO. The obtained curves are the best fit to the intensity of peaks A and B of the $L_3$ edge, as indicated by dashed line.

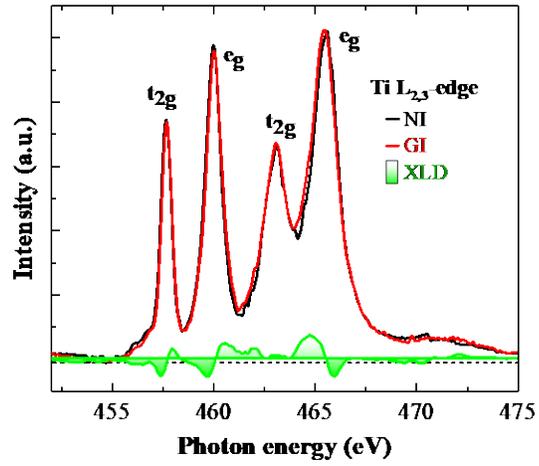

**Figure S5. Characterizing the Ti Valence state.** X-ray absorption spectroscopy (XAS) characterization result for *c*-LAO/1 uc LFO/STO at Ti L (2p-3d transition) edge with normal incident angle (NI) $\Theta = 80°$ and grazing incident angle (GI) $\Theta = 10°$. Here the signal of XLD in green color is defined as XLD = (NI-GI), indicates the orbital occupancy.

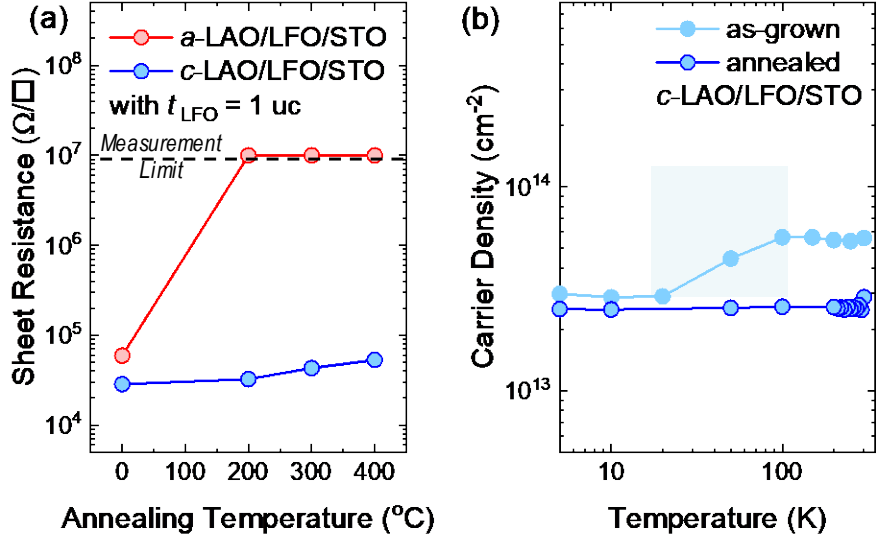

**Figure S6. Thermal treatment of amorphous and crystalline samples.** (a) Sheet Resistance as a function of annealing temperature in *a*- and *c*-LAO/LFO/STO samples with $t_{LFO} = 1$ uc. (b) Carrier densities as a function of temperature in as-grown and annealed *c*-LAO/LFO/STO samples.

The transport properties of post-annealed crystalline and amorphous LaAlO$_3$/LaFeO$_3$/SrTiO$_3$ ($t_{LFO} = 1$ uc) are summarized in Figure S6. As clearly shown in Figure S6(a), thermal annealing at a temperature ~200 °C in air for 1 hour can effectively remove oxygen vacancies and thus makes the *amorphous* interface completely insulating. Here, a relatively low annealing temperature is chosen to avoid an interfacial intermixing and crystallization of the amorphous LaAlO$_3$ layer during the annealing. On the other hand, the conducting interface of the crystalline LaAlO$_3$/LaFeO$_3$/SrTiO$_3$ (*c*-LAO/LFO/STO) remains conducting after such a post-annealing. In addition, as shown in Figure S6(b), the *annealed c*-LAO/LFO/STO sample shows nearly degenerate carrier density, while the as-grown sample exhibits carrier localization around 20-100 K due to oxygen vacancy carriers.[2] This observation suggests that the redox reaction (with oxygen vacancies) should be the only mechanism for the interlayer charge transfer in the amorphous samples, while both the redox reaction and polar discontinuity (with charges surviving from annealing) contribute to the conducting crystalline interface.

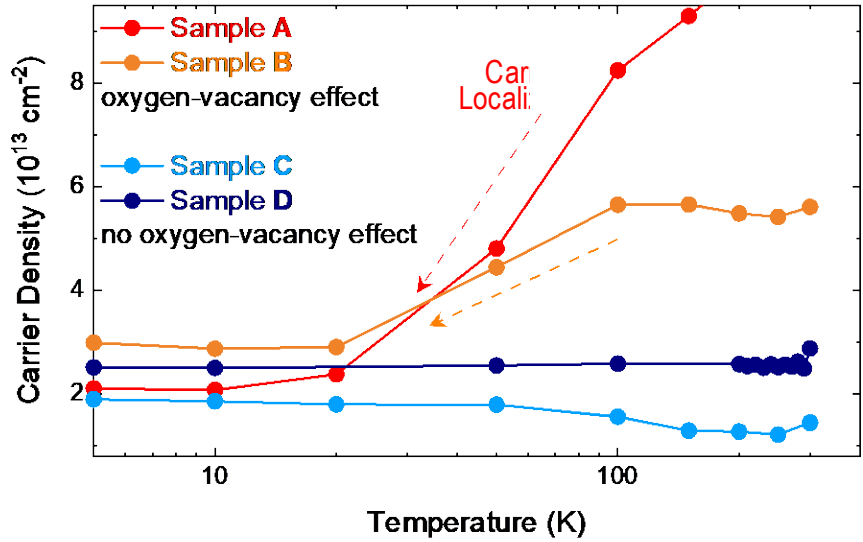

**Figure S7**. Temperature-dependent carrier densities obtained from Samples **A**–**D**. The carrier reduction around 20-100 K is observed in Samples **A** and **B**, indicating a dominant role of oxygen vacancies in electrical transport measurement. On the other hand, carrier densities are less temperature-dependent in Samples **C** and **D**.

Assuming there are some oxygen vacancies in STO, trapping states will be created below the Fermi level and result in the carrier localization around 20-100 K. Figure S7 compares the temperature-dependent carrier densities $n_S$ of the as-grown *a*-LAO/STO (sample **A**), as-grown *c*-LAO/LFO/STO ($t_{LFO}$ = 1 and 2 uc, Samples **B** and **C**), and annealed *c*-LAO/LFO/STO ($t_{LFO}$ = 1 uc, Sample **D**). While the samples **A** and **B** show a clear drop of $n_S$ (or carrier localization) below 100 K, the samples **C** and **D** exhibit less-temperature-dependent $n_S$ at low temperatures. This phenomenon reveals the transport property of samples **A** and **B** is affected by oxygen vacancies, while the effect of oxygen vacancies is negligible in samples **C** and **D**. So, the formation of oxygen vacancies is suppressed by either post-annealing or increasing $t_{LFO}$.

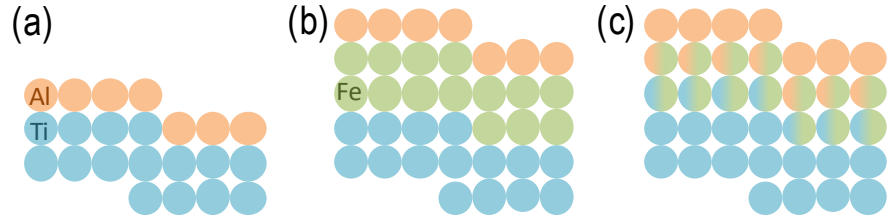

**Figure S8**. Sketch of atomic contact at the LFO-buffered LAO/STO interface, and only the B-site ions (Al, Ti and Fe) are plotted for convenience. (a) Standard LAO/STO interface with a direct Ti-O-Al bonding, or Ti-Al contact in the simplified sketch. (b) In an ideal heterostructure, the 2-uc LFO buffer layer completely blocks the Ti-Al contact. (c) In a real case with 1 uc intermixing layer of Al-Fe and Ti-Fe, the 2-uc LFO buffer layer can block most of, but not all the Ti-Al contact.

The atomic contact between LAO and STO (i.e, Ti-O-Al bonding or Ti-Al contact in a simplified sketch in Figure S8) is required for the redox reaction at the LAO/STO interface, as shown in Figure S8(a). If inserting an inert buffer layer (like $LaMnO_3$, $EuTiO_3$ or LFO) that is thick enough to block such Al-Ti contact in Figure S8(b), the redox reaction between LAO and STO will be avoided. However, the 1 uc buffer layer is usually not thick enough. Given the atomic steps and interfacial intermixing (1 uc layer of Ti-Fe and Fe-Al intermixing) at a real heterointerface as sketched in Figure S8(c), a 2 (3) uc thick LFO layer will block most (all) of the Ti-O-Al bonding to induce a metal-insulator transition at the buffered LAO/STO interface. This is also consistent with experimental results in previous studies on the buffered-LAO/STO systems,[5] as well as our results in the LFO-buffered LAO/STO. So, we believe this 2-3 uc is a universal characteristic length of redox-reaction-mediated interlayer charge transfer.

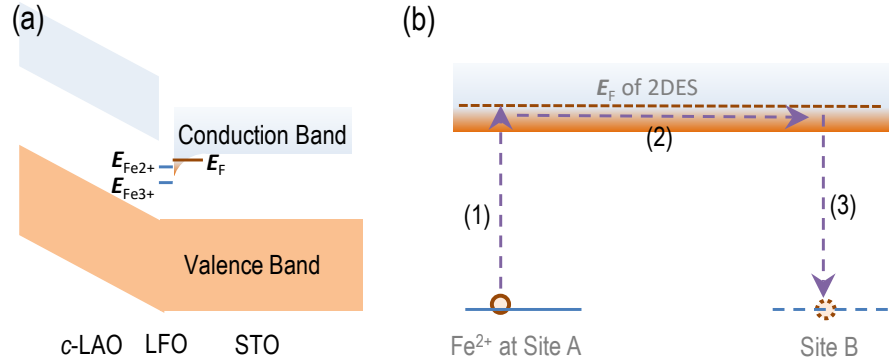

**Figure S9**. (a) Sketch of band structure at *c*-LAO/LFO/STO with $t_{LFO}$ = 4 and 5 uc. (b) Three steps, indexed by (1), (2) and (3), for the charge movement in the LFO layer.

To check the validity of our estimated LFO carrier density $n_{LFO}$ ($n_{LFO} \sim n_C = 3.5 \times 10^{17}$ cm$^{-3}$), we build another physic model to explain the charge movement in the LFO layer, of which the origin of mobile charges is ascribed to the electron hopping between LFO and STO. During the interlayer charge transfer from both polar catastrophe and redox reaction, the LFO buffer layer always serves as an energetically favored electron acceptor due to the small bandgap and low oxygen affinity. This is confirmed by the observation of B-site $Fe^{2+}$ ions in EELS and XAS data, corresponding to the $Fe^{3+}$ ions receiving transferred electrons at the B-site. Moreover, for the 4-5 uc LFO-buffered *c*-LAO/STO sample, the energy-level position of $Fe^{2+}$ ($E_{Fe2+}$, one electron occupies the $Fe^{3+}$ state at B-site) should be close to (or slightly below) the Fermi level $E_F$ of 2DES in STO due to the coexisting $Ti^{3+}$ and $Fe^{2+}$, as sketched in Figure S9(a). This is because if the $Fe^{2+}$ level is higher (or much lower) than $E_F$, there should be no $Fe^{2+}$ in LFO (or no 2DES in STO). Therefore, the charge movement in LFO layer from Site A to B can be divided into three steps: 1) electron excitation from $Fe^{2+}$ to 2DES at the site A with overcoming the energy barrier $E_{gap}$, 2) electron movement along the STO conduction band to site B, and 3) electron falling back to Fe to form $Fe^{2+}$ at Site B, as plotted in Figure S9(b). In this case, $n_{LFO}$ can be described by $n_{LFO} = n_{Fe2+} \exp(-E_{gap}/k_B T)$, where $n_{Fe2+}$ is the $Fe^{2+}$ density in LFO layer, $E_{gap}$ is the energy difference between $E_{Fe2+}$ and $E_F$, $k_B$ is the Boltzmann constant and T is the temperature.

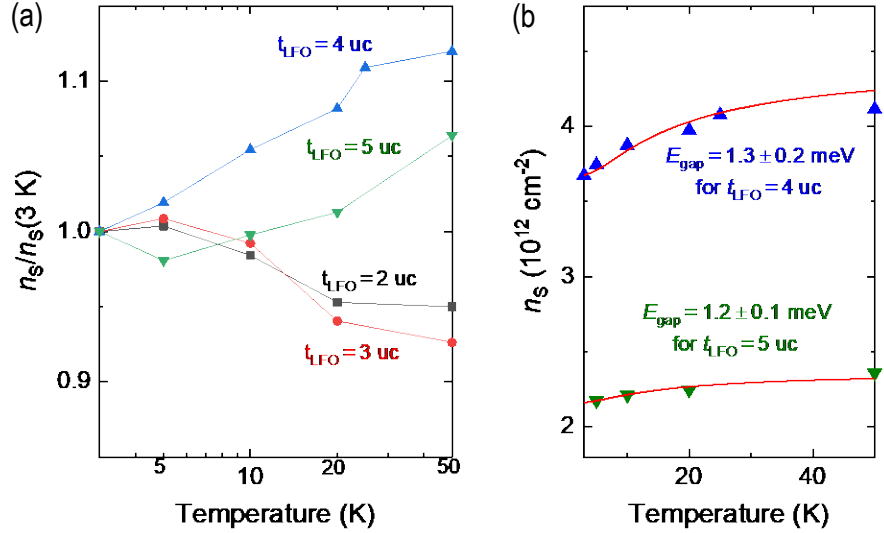

**Figure S10**. (a) Comparison between normalized $n_S$ at samples with different $t_{LFO}$. (b) the fitting curves (red line) and experiment data (scatters) for $t_{LFO}$ = 4 and 5 uc samples. The estimated Egap is around 1.2–1.3 meV.

In the equation $n_{LFO} = n_{Fe2+} \exp(-E_{gap}/k_BT)$, $n_{Fe2+}$ can be estimated by EELS and/or XAS data shown in Figure 3 in main text and Table S1 in supporting information, with $n_{Fe2+} \sim 6\text{–}7 \times 10^{20}$ cm$^{-3}$ for $t_{LFO}$ = 4-5 uc. The value of $E_{gap}$ can be estimated by the slight drop of $n_S$ in 2DES on cooling at low temperatures (3-50 K) in $t_{LFO}$ = 4-5 samples, with $n_S = n_{polar} + n_{Fe\text{-}Ti}$. Here, $n_{polar}$ is due to the screened polar LAO and less depending on temperature, while $n_{Fe\text{-}Ti}$ is temperature-dependent and corresponds to the charge density excited from Fe$^{2+}$ to 2DES with $n_{Fe\text{-}Ti} \propto \exp(-E_{gap}/k_BT)$. If the screening effect of the LFO buffer layer is weak with less Fe$^{2+}$ ions, such $n_S$ drop will not be observable. This is supported by Figure S10(a), where the thinner LFO layer ($t_{LFO}$ = 2-3 uc) with the weaker screening effect doesn't induce the carrier drop below 50 K. In Figure S10(b), we show the fitting of low-temperature $n_S$ with $E_{gap} \sim 1.2\text{-}1.3$ meV for the $t_{LFO}$ = 4-5 uc. So, the value of $n_{LFO}$ at 2 K can be calculated around using $n_{LFO} = n_{Fe2+} \exp(-E_{gap}/k_BT)$ with $n_{Fe2+} = 6\text{–}7 \times 10^{20}$ cm$^{-}$ [3] and $E_{\text{gap}}$ = 1.2-1.3 meV. The calculated $n_{\text{LFO}}$ is around 4–5 × 10$^{17}$ cm$^{-3}$, which is reasonably consistent with $n_{\text{C}}$ ~ 3.5 × 10$^{17}$ cm$^{-3}$. Also, if using $n_{\text{LFO}}$ ~ 4–5 × 10$^{17}$ cm$^{-3}$ to calculate screening length, the result is 5.1–5.6 uc, which is still consistent with our observation that the polar-discontinuity-induced interlayer charge transfer between LAO and STO is screened by a 6-uc-thick LFO buffer layer in LAO/LFO/STO heterostructure.

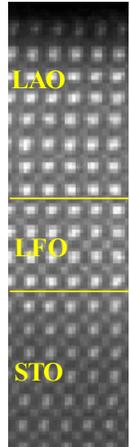

| Fe Valence fitting | Intensity $L_3/L_2$ | $Fe^{2+}$ % |
|---|---|---|
| $Fe^{3+}$ reference | 5.16 | |
| $Fe^{2+}$ reference | 3.48 | |
| 1 Fe | 4.75 | 24.17 % |
| 2 Fe | 5.15 | 0.76 % |
| 3 Fe | 5.16 | 0.42 % |
| 4 Fe | 5.16 | 0.96 % |

**Table S1.** Fe-$L_{2,3}$ edge peak fitting obtained from the 4-uc-thick LFO buffer layer. After fitting the Fe-$L_{2,3}$ edge peak by a combination of Gaussian and Lorentzian functions, the ratio between intensities of $L_3$ and $L_2$ ($L_3/L_2$) was performed to calculate the $Fe^{2+}$ concentration, denoted by $Fe^{2+}/(Fe^{2+}+Fe^{3+})$. From Table S1, the $Fe^{2+}$ fraction is 24±5% in the top LFO plane and reduces close to 0% for the rest of the LFO planes.